# The association between Marital Locus of Control and break-up intentions

David Boto-García*

botodavid@uniovi.es
Department of Economics
University of Oviedo

Federico Perali

federico.perali@univr.it
Department of Economics
University of Verona

**Abstract:**

Understanding couple instability is a topic of social and economic relevance. This paper investigates how the risk of dissolution relates to efforts to solve disagreements. We study whether the prevalence of relationship instability in the past among couples is associated with marital locus of control. This is a noncognitive trait that captures individual's perception of control over problems within the couple. We implement a list experiment using the count-item technique to a sample of current real-life couples to elicit truthful answers about couple break-up intentions in the past at the individual level. We find that around 44% of our sample has considered to end their relationship with their partner in the past. The intention to break-up is more prevalent among those who score low in marital locus of control, males, low-income earners, individuals with university studies and couples without children.

**JEL codes:** J12, C91

**Keywords:** *couple break-up, marital locus of control, count item technique; list experiment*

**Conflict of interest:** The author(s) have no conflict of interest to report.

**Acknowledgements:** We acknowledge useful comments and suggestions from Martina Menon and Antonio Alvarez. The authors also wish to thank financial support by the Spanish Ministry of Education, Culture and Sport (FPU 16/00031).

**Declaration of interests:** NONE

*Corresponding author
Avenida del Cristo, 33006; Faculty of Economics and Business, University of Oviedo, Asturias (Spain)

## 1. Introduction

After the seminal contributions of Becker (1973; 1974), scholars in economics and sociology have devoted increasing attention to the study of couple stability and its determinants due to its important economic effects on fertility and childrearing (Chiappori & Weiss, 2006; Lundberg et al., 2016), children's engagement into risky behaviours (Gustavsen et al., 2016), and children's lower earnings during adulthood through reduced happiness (Mohanty & Ullah, 2012) and lower investments in human capital (Garasky, 1995), among others. Empirical evidence has shown that the likelihood of couple break-up is positively associated with job instability (Ahituv & Lerman, 2011), workplace contact between opposite sexes (McKinnish, 2004), low emotional stability (Lundberg, 2012) and husbands' paid leave for childrearing (Avdic & Karimi, 2018), among many other factors.

Little attention has been paid to spouses' *intentions* to separate as a result of a couple crisis.[1] Economic models of marriage and divorce suggest that the decision to end a sentimental relationship is driven by the expected gain of remaining together compared to becoming single (Becker et al., 1977; Pollak, 2019). This seems to vary across individuals' personal characteristics, with most break-ups being unilateral. Rather than studying the determinants of break-ups ex post, in this study we are interested in exploring truthfully revealed partners' *intentions* to finish their sentimental relationships. This is crucial information to design and implement effective prevention policies that could alleviate intra-couple problems and dissatisfaction.

Most empirical studies use data on actual divorce or legal separation in married couples (Killewald, 2016; Zulkarnain & Korenman, 2019). However, in the last decades, there has been an increase in the prevalence of cohabitation among young generations (Eickmeyer & Manning, 2018; Rosenfeld & Roesler, 2019) and the consequent delay in marriage and childrearing. This behaviour is probably due to the lower gains of marriage from the gender division of labor (Stevenson & Wolfers, 2007), the availability of birth control methods and the legalization of abortion (Choo & Siow, 2006), declining social stigma against cohabitation and premarital sex (Rosenfeld & Roesler, 2019), and because of alternative relationships being more readily available (McKinnish, 2004). Because cohabiting couples are generally less satisfied than married ones (Dilmaghani, 2019) and cohabitation imposes less barriers to couple dissolution because of reduced transaction costs, it seems that recently formed couples without a formal marriage contract might be more exposed to the risk of dissolution.

[1] Throughout the paper we make use of the terms 'spouses' and 'partners' interchangeably to refer to each member of the couple, no matter whether they are married or not.

In this study, we examine partners' past intentions to end their sentimental relationships using data from a sample of current married and non-married couples collected in an experimental setting. We analyse the prevalence of past intentions to leave the partner and how this correlates with individual characteristics. We are especially interested in exploring whether couple break-up intentions in the past relate to marital locus of control (MLC). MLC is a psychological trait that measures partners' view about whether problems within the couple are contingent on their own efforts or due to external factors (Miller et al., 1983). From a theoretical perspective, MLC can be seen as an indicator of adjustment costs under frictions that determines couple stability through better capacity of making binding commitments (Lundberg & Pollak, 2007). Accordingly, unlike other studies that relate overall Locus of Control with couple satisfaction (e.g., Lee & McKinnish, 2019), we consider locus of control defined in the marital domain. In our analysis, spouses who score low in MLC attribute couple problems to external circumstances beyond their control such as luck or fate, thereby feeling less responsible for the relationship problems.

We employ survey data collected in a lab experiment conducted in four cities in the north of Spain involving 266 individuals (133 couples). Because direct self-reports about past break-up intentions could be heavily affected by social desirability bias (Tourangeau & Yan, 2007), we conduct a list experiment in the form of a count-item technique that allows us to elicit such a sensitive question truthfully.[2] This indirect questioning technique is often used to elicit truthful answers to sensitive questions and has been successfully applied for uncovering criminal behaviour (Kuha & Jackson, 2014), political voting (Brownback & Novotny, 2018), racial prejudice (Imai, 2011), or attitudes towards female genital cutting (Cao et al., 2018). Because we work with a cross-sectional dataset, it is important to make it clear that we cannot give our estimates a causal interpretation since the endowment of marital locus of control could be correlated with other psychological traits. Even though auxiliary checks do not detect problems of endogeneity, we interpret our results in terms of associations, as done by Lee and McKinnish

[2] Social desirability bias refers to the tendency of people to answer questions that are perceived as sensitive or private in the way they believe the society considers as appropriate. It generally makes people to overreport 'good behaviour' and underreport 'bad behaviour'.

(2019). Our experimental setting is therefore performed to identify a sensitive issue (past intentions to leave the current partner) truthfully due to social desirability bias, but we cannot manipulate marital locus of control, which we take as a given psychological trait.

The paper contributes to the economic and sociological literatures on couple instability by shedding light on the role of problem-solving attitudes in conflicts resolution within the couple. Based on a linear difference-in-mean estimator with interactions, we show that the share of individuals who have seriously considered to break with their current partner in the past is about 44%. Past couple crises are found to be negatively associated with MLC, implying that individuals who devote low effort to couple problem solving and consider the relationship matters to be beyond their control are at higher risk of couple dissolution. To the best of our knowledge, this is the first study that addresses past attempts to separate among couples that are currently together and relates strains in the relationship to the marital locus of control trait.

By studying past break-up intentions in a sample of current couples, one could argue that, aside from sampling issues, the target population suffers from self-selection (as they have managed to continue together). We want to make it clear from the scratch that *current* couples are indeed the population of interest here.[3] Rather than studying the drivers of actual break-ups (Killewald, 2016; Stevenson & Wolfers, 2007; Zulkarnain & Korenman, 2019), we look at the correlates of past *intentions* to break-up. Couple problems are important predictors of later actual split in the future (Amato & Rogers, 1997). In our view, this is highly informative about the factors that make couples at risk of dissolution but without actual split and/or that helps couples to overcome relationship crises. Some couples continue together because of low exit possibilities and the emotional costs associated with separating even if the relationship quality is not as desired (Lundberg & Pollak, 1996), living in a sort of 'internal divorce' (Konrad and Kommerud, 2000) and 'separate spheres' (Lundberg & Pollak, 1993).

The remainder of the paper is structured as follows. Section 2 presents the theoretical framework for the analysis. Section 3 describes the experimental setting, survey design, data and variables used. Section 4 outlines the list experiment. Section 5 presents and discusses our findings, together with some endogeneity checks. Finally, Section 6 summarizes the results and concludes with policy considerations.

[3] By construction, any study of couple behavior is affected by individuals' self-selection into partnerships.

## 2. Theoretical background and related literature

The theoretical background for the economic analysis of couple instability dates back to Becker (1973; 1974) and Becker et al. (1977). Individuals marry or engage in a sentimental relationship when the expected utility from being together exceeds the expected utility from remaining single or finding another partner, which depends on the efficiency of the alternative match. Marital sorting is therefore the result of a search process in which individuals draw a potential partner from a distribution of candidates. Search is costly and depends on age and the sex ratio in the (re)marriage market (Chiappori et al., 2002; Abramitzky et al., 2011).

Due to information asymmetries and uncertainty, finding the optimal match might involve large search costs. In case of a minimum acceptable draw, individuals must decide whether to continue searching for the optimal sorting or to accept a potential degree of mismatch and engage in a relationship. The expected discounted utility of the relationship is initially determined by the imperfect information individuals have at that moment about themselves and about the potential partner (Pollak, 2019). To reduce this uncertainty, individuals match with similar mates based on both observed and unobserved characteristics. In this regard, a large body of research documents assortative matching based on education, age or income (Choo & Siow, 2006; Chiappori et al., 2018). This is argued to be due to gains from specialization (Pollak, 2012) and because frictions are more likely to emerge between mates with opposite characteristics. Recent research shows that assortative matching is also based on personality traits and cognitive ability (Dupuy & Galichon, 2014; Flinn et al., 2018).

Bergstrom (1996) maintains that at the beginning of a partnership it is not possible to write a pre-relationship contract that legally binds the partners to a detailed program of behaviour through the course of the relationship. Investing on a stable relationship requires to solve disagreements through patient negotiations (bargaining). In the absence of a household head and following traditional cooperative and bargaining models of household decision-making (Manser & Brown, 1980; McElroy & Horney, 1981; McElroy, 1990), partners are players of repeated games that maximize their utilities subject to the other's preferences and their bargaining weights (Chen & Woolley, 2001, Chavas et al., 2018). Assuming that partners freely communicate and can make binding self-enforcing agreements, cooperative game theory assumes that the equilibrium outcomes are Pareto optimal. Intra-household bargaining requires both partners to discuss matters and concede to achieve the intended Pareto efficient allocations (Zeuthen, 1930). In this vein, the sustainability of efficient solutions is heavily dependent on each individual's attitudes (Andaluz & Molina, 2007). The perception about the capability of discussing and fixing matters and the willingness to adopt relationship-maintenance strategies when couple problems arise might produce a great impact in the functioning of the relationship (Canzi et al., 2019; Pagani et al., 2019). The locus of control personality trait, applied to the couple sphere, has been shown to explain the capacity of couples to solve their disagreements (Miller et al., 1986).

Based on Becker's framework (Becker et al., 1977), dissolution is a stochastic event that depends on the expected gains from continuing in the relationship and the variance of their distribution. Both the mean and the variance of the discounted gains are affected by the anticipated capacity of solving unexpected problems. Once the noisy initial signal about the capacity to get along with each other is updated, individuals adjust the utility of continuing with the relationship relative to quitting. Deluded expectations are often responsible for separations. If partners further consider their couple problems are beyond their control and place low effort on negotiation and concession, this makes credible long-term binding commitments unsustainable, which in turn increases couple instability (Lundberg & Pollak, 2007). As such, marital locus of control can be seen as an indicator of adjustment costs. The lower the effort to compromise, the higher the probability of frictions among partners. As such, we set the following research hypothesis:

H0: Break-up intentions are negatively associated with marital locus of control.

The psychological and sociological literatures have shown that those with internal locus of control are better problem solvers and are more satisfied with their partner (Doherty, 1981; Myers & Booth, 1999). Presumably, this is because they are more willing to invest the time and energy required to keep their sentimental relationship healthy. Using longitudinal data, Kubacka et al. (2011) and Donato et al. (2015) report that increases in dyadic coping responses increase relationship satisfaction due to the encouragement of feelings of gratitude and reciprocity. Recently, Lee and McKinnish (2019) have shown that higher scores in internal locus of control are associated with higher marital satisfaction. Interestingly, they also show that own locus of control trait is more relevant than the partner's locus of control and that spouses with low scores in marital locus of control report declines in marital satisfaction over time.

A key point in this framework is that although the dissolution of a couple involves both partners, the decision to end the relationship is taken individually. The valuation of the net utility from remaining together may differ across partners based on their characteristics and expectations. Accordingly, our analysis is performed at the individual level.

Not all couples that are unable to achieve Pareto efficient outcomes split up. Much depends on the expectations each partner has about the probability to improve the quality of the match in the remarriage market. Both the level and volatility of income also plays a key role for partners' *voice* and *exit* possibilities (Nunley & Seals, 2010), especially for females (Gonalons-Pons & Calnitsky, 2022; Parkman, 2004). Further, some couples prefer to remain together because there are important emotional costs associated with separating. This relates to what Konrad and Lommerud (2000) term 'internal divorce'. That is, separation without leaving the household so that the members of the couple live in 'separate spheres' (Lundberg & Pollak, 1993). In this vein, Chavas et al. (2021) show that in most cases agreements are not efficient but sufficiently *satisficing*. Social norms and the legal framework for children custody may also contribute to prevent a couple from entering a formal separation process. This is because the anticipated gains from dissolution are lower than the personal losses associated with continuing together, even if the quality of the relationship is not as desired (Lundberg & Pollak, 1996). In any case, although past couple break-up intentions did not materialize into formal separation, they are good predictors of intra-couple instability, dissatisfaction and future divorce (Guven et al., 2012) and therefore require further attention.

## 3. Data

### *3.1. Experimental setting and survey design*

We conducted an artefactual lab experiment on a sample of couples in four northern Spanish cities (Oviedo, Gijon, Avilés and Bilbao). We performed several lab sessions from June to November 2019. Participation was voluntary and participants were sampled from the general population through advertisements and posters. The only requirement was to be in a sentimental relationship and to be older than 18. In the announcements, it was clearly stated that the participation of both partners was required. A total of 133 real-life couples (266 individuals) took part in the experiment. Out of them, only three were same-sex couples. Since they represent a low share, a separate analysis of heterosexual and homosexual couples is not feasible so these three couples were disregarded.

In each session, we conducted both a Public Goods Game and a Discrete Choice Experiment for a holiday trip, which are not analysed here.[4] At the end of each session, participants answered an individual questionnaire (see Sections A and B in Online Appendix). The whole experiment took about an hour. Each participant received a show-up fee of 10€ plus some additional earnings based on the allocations made in the Public Goods Game. In this way, participation in the experiment was monetary incentivized. Each participant was paid in cash individually at the end of the experiment. Each couple earned €31 on average.

Prior to the beginning of the experiment and the survey, participants were instructed that the study is only for research purposes and were guaranteed that their answers would be kept private. Participants were assigned an ID code and spouses within couples were separated in two different rooms during the tasks. Participants were seated at a distance guaranteeing sufficient privacy during the completion of the tasks and the survey. Instructions were read aloud and handed in paper at the beginning of each task.

### *3.2. Marital locus of control (MLC)*

We implemented a reduced version of the 44-item Miller's marital locus of control scale by selecting eight statements from Miller et al. (1983). Individuals were asked to report their degree of agreement with eight statements (see Table 1) on a 1-7 Likert scale, where 1 means 'I totally disagree' and 7 means 'I totally agree'. These statements aim to capture the extent to which one regards the couple wellbeing and stability as being under one's effort and control (statements 2, 3, 5 and 7) versus being chance-determined, incidental, and beyond partner's control (statements 1, 4 and 6). To make them comparable, answers to statements 1, 4 and 6 were reversed so that higher values indicate higher MLC.

[4] The general aim of the study was to understand intra-household bargaining in two different domains: (i) a Public Goods Game using the voluntary contribution mechanism to household public goods, and (ii) a Choice Experiment related to the choice of a vacation destination and the potential conflict associated with the decision. The order of the two experiments was randomized.

**Table 1.** Statements used for measuring MLC

| Question: | Please indicate your agreement/disagreement with the following statements on a 1-7 scale where 1 means "I STRONGLY DISAGREE" and 7 means "I STRONGLY AGREE": |
|---|---|
| Statement number | |
| 1 | *It is unavoidable that throughout a relationship some conflicts and anger emerge, no matter whether one do things as good as possible.* |
| 2 | *When my partner and I got angry with each other for any reason, I have the ability to become reconciled easily.* |
| 3 | *When my partner and I got angry with each other for any reason, I am the person who moves first to solve the conflict* |
| 4 | *When my partner and I have an argument or disagreement, we let time to pass until we forget about it.* |
| 5 | *Success in a relationship merely depends on the effort each one puts into it. If both strive, the relationship works.* |
| 6 | *My partner's moods are often mysterious to me, in that I have little idea to what may have set them off.* |
| 7 | *I believe my partner and I would remain together and happy even under bad and extreme situations.* |
| 8 | *I consider that the success of a relationship depends on clear communication. Honesty and discussing problems and thoughts with your partner are crucial factors for a successful relationship.* |

Source: adapted from Miller et al. (1983).

Table 2 presents descriptive statistics of the answers to the eight statements, where *r* indicates that the statement is reversed coded. Respondents moderately agree, on average, with statements 1, 7 and 8. Concerning the other statements, opinions are more balanced, especially for statements 4 and 6 that exhibit the largest standard deviations.

**Table 2.** Descriptive statistics of the MLC statements (N=266)

| Statement | Mean | SD | Min | Max |
|---|---|---|---|---|
| 1r | 1.729 | 1.189 | 0 | 7 |
| 2 | 4.992 | 1.443 | 1 | 7 |
| 3 | 4.444 | 1.448 | 1 | 7 |
| 4r | 4.481 | 2.364 | 0 | 7 |
| 5 | 5.000 | 1.709 | 1 | 7 |
| 6r | 4.974 | 2.204 | 0 | 7 |
| 7 | 6.034 | 1.022 | 2 | 7 |
| 8 | 6.545 | 0.777 | 3 | 7 |

In line with Lee and McKennish (2019), we construct an aggregate indicator of MLC by summing each respondents' answers to the eight statements. Higher values of this score thus proxy higher internal MLC. Since there are no *a priori* reasons to assign different weights to each statement, the index construction assumes equal weights.[5] This variable ranges from 23 to 50, with a mean equal to 38.2. To better interpret this indicator, the variable is standardized to have zero mean and standard deviation equal to one (denoted by *marlocus*). A kernel density plot for the original and the standardized variable is presented in Figures A1 and A2 in the Online Appendix.

### *3.3. Descriptive statistics*

In the questionnaire, we collected participants' sociodemographic characteristics such as age, gender, education level, net monthly individual income, labour status, health status, nationality, the length of the relationship and marriage status, among others. Furthermore, they were also asked to rate on a 0-10 Likert scale their patience (where 0 means "Nothing at all" and 10 means "A great deal") and willingness to take risks in general (where 0 means "Nothing at all" and 10 means "A great deal"). We used the same wording as in the German SocioEconomic Panel. Visher et al. (2013) and Dohmen et al. (2011) have validated these ultra-short measures of patience and willingness to take risks.

Table 3 reports the descriptive statistics of the sample. Young people and those with university studies are slightly overrepresented in our sample, so this potential sample selection should be bear in mind when interpreting the findings. This is a common feature in experimental studies in which participation is voluntary (Levitt & List, 2007). Average age is 32 years old and 61% attain university studies. In terms of income and labour status, the sample is more balanced, although 32% of respondents are students. Most participants are Spanish (97%) and are in good health conditions (96%). Around 29% of couples are married, with an average number of children of 0.43, ranging from 0 to 3. About half of the sample (53%) have been in a relationship for less than 5 years, with 15% staying together for more than 25 years. Individuals are moderately patient, with an average score equal to 6.3. Concerning the willingness to take risks, respondents locate themselves in the midpoint of the scale. However, the standard deviations suggest relevant variability in these traits (see Figures A3 and A4 in the Online Appendix for the frequencies).

[5] Nonetheless, we construct an alternative index based on Principal Component Analysis as a robustness check (see subsection 5.3).

**Table 3.** Descriptive statistics (N=266)

| **Variable** | **Description** | **Mean/ %** | **SD** | **Min** | **Max** |
|---|---|---|---|---|---|
| Individual characteristics | | | | | |
| *age* | Age in years | 32.74 | 14.16 | 18 | 89 |
| *female* | =1 if female | 50.75 | | | |
| *prim.educ* | =1 if primary education | 7.89 | | | |
| *sec. educ* | =1 if secondary education | 30.82 | | | |
| *univ.educ* | =1 if high education | 61.27 | | | |
| *income0* | =1 if NMII=0 | 27.44 | | | |
| *income1* | =1 if NMII <€500 | 13.53 | | | |
| *income2* | =1 if NMII between €500 and €1,500 | 29.69 | | | |
| *income3* | =1 if NMII between €1,500 and €2,500 | 22.93 | | | |
| *income4* | =1 if NMII >€2,500 | 5.26 | | | |
| *income* | =0 if NMII=0; =1 if 0<NMII≤€500; =2 if €500<NMII≤€1,500; =3 if €1,500<NMII≤€2,500; =4 if NMII>€2,500 | 1.628 | 1.259 | 0 | 4 |
| *employed* | =1 if currently employed | 54.51 | | | |
| *unempl* | =1 if unemployed | 6.01 | | | |
| *inactive* | =1 if inactive (retired/disabled/housekeeper) | 7.51 | | | |
| *student* | =1 if student | 31.95 | | | |
| *health* | =1 if respondent is in good health conditions | 95.86 | | | |
| *natspain* | =1 if respondent is Spanish | 97.74 | | | |
| *marlocus* | Marital locus of control scale (sum of statements 1-8, standardized) | 0.00 | 1.00 | 0 | 1 |
| *patience* | 0-10 Likert scale of self-assessed patience | 6.33 | 2.57 | 0 | 10 |
| *risk* | 0-10 Likert scale of willingness to take risks | 5.62 | 2.59 | 0 | 10 |
| Couple characteristics | | | | | |
| *married* | =1 if married | 29.32 | | | |
| *numchildren* | Number of children | 0.436 | 0.780 | 0 | 3 |
| *rel_lessfive* | =1 if respondent is in a relationship for less than 5 years | 53.00 | | | |
| *rel_fivetofifteen* | =1 if respondent is in a relationship for more than 5 years but less than 15 | 24.43 | | | |
| *rel_fifteentotwfive* | =1 if respondent is in a relationship for more than 15 years but less than 25 | 15.03 | | | |
| *rel_moretwfive* | =1 if respondent is in a relationship for more than 25 years | 43.60 | | | |

*Note: NMI means net monthly individual income

Table 4 reports a descriptive OLS regression of MLC on respondents' characteristics reveals that MLC trait is positively associated with the willingness to take risks and inversely associated with age (at 10% significance level). Besides this, females appear to exhibit a greater MLC. In contrast, it is unrelated to patience, education level, the tenure of children or how long the couple has been together. With regard to the link between partners' individual MLC, Figure 1 presents a scatter plot of the association between the two. We see there is a positive and statistically significant correlation (corr=0.324), implying that individuals match with partners with similar MLC trait. This falls in line with a large literature on assortative matching (Dupuy & Galichon, 2014; Flinn et al., 2018).

**Table 4.** Descriptive OLS regression of *marlocus* on personal characteristics

| Dep. Variable: *marlocus* | Coefficient (SE) |
|---|---|
| *age* | -0.004* |
| | (0.002) |
| *female* | 0.044* |
| | (0.025) |
| *Sec.educ* | -0.008 |
| | (0.050) |
| *univ.educ* | -0.035 |
| | (0.048) |
| *Length relationship: less 5 years* | 0.069 |
| | (0.075) |
| *Length relationship: 5-15 years* | -0.006 |
| | (0.063) |
| *Length relationship: 15-25 years* | 0.075 |
| | (0.057) |
| *children* | 0.079 |
| | (0.048) |
| *patience* | 0.006 |
| | (0.005) |
| *risk* | 0.011** |
| | (0.005) |
| constant | 0.582*** |
| | (0.139) |
| Observations | 266 |

Standard errors in parentheses. *** p<0.01, ** p<0.05, * p<0.1

**Figure 1.** Scatter plot of the association between males' and females' MLC

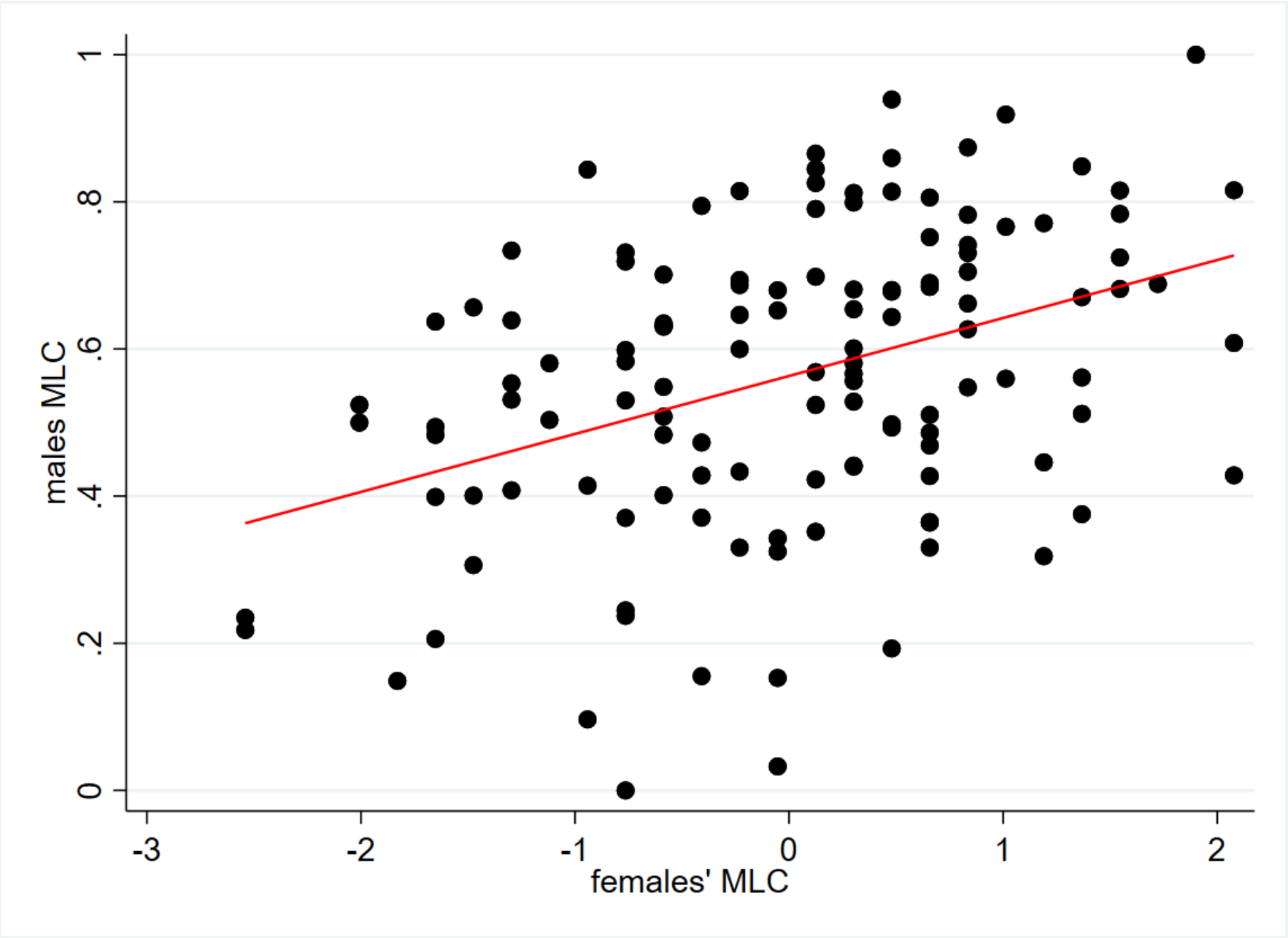

## 4. The Count Item Technique

When asked directly about sensitive topics such as norm violations or intimacy, individuals tend to misreport their behaviour: they are reluctant to admit publicly some attitudes. In our setting, several works have documented that people underreport infidelity when asked directly (Munsch, 2015; Tang et al., 2022). The same is likely to apply is they are asked directly about their past intentions to end their current sentimental relationship. To overcome this limitation, Miller (1984) introduced the count-item technique, also known as 'the list experiment'. This technique consists of randomizing a survey sample into two groups. One of them receives a list of $J$ items (statements) and the other receives the same list plus an additional item ($J + 1$) that addresses the sensitive question. In both cases, to ensure the confidentiality of responses to the sensitive issue, respondents are asked the number of items to which the individual agrees, or which applies to her (see below).

In our case, half of the sample (hereafter CONTROL group) was prompted with the following question:

*Now I am going to enumerate you a list of three issues that might or might not be true for you. Please indicate just the NUMBER of them (0,1,2,3) that are correct in your case. Do not tell me which of them but how many:*

- *I have travelled by car without the seat belt at least once (in the last month)*
- *I have celebrated a meeting with my family or some friends (in the last month)*
- *I have been abducted by aliens*

*The number of statements that are true for me are:* ________

The other half (hereafter TREATED group) was presented with the same question but with the following additional statement (henceforth the key item):

- *I have been about to seriously end my relationship with my partner/spouse*

In each case, individuals are asked to indicate the number of statements that are true in their case, thereby allowing them to encrypt their answer to the sensitive one. Importantly, the proper implementation of the count-item technique relies on the following three assumptions:

- *Treatment randomization*: the assignment of respondents to the TREATED and the CONTROL groups is merely random.
- *No design effect*: answers to the non-key statements are not different between the CONTROL and the TREATED group.
- *No liar*: respondents are assumed to answer the sensitive issue in the list experiment truthfully.

Another relevant issue concerns the definition of the non-sensitive statements. They are usually selected in a way that avoids, on average, extreme answers (named ceiling and floor effects). Glynn (2013) recommends defining statements that are negatively correlated. Following this advice, we chose a low-prevalent item (*I have travelled by car without the seat belt at least once in the last month*) and a high-prevalent one (*I have celebrated a meeting with my family or some friends in the last month*) and a highly unlike one (*I have been abducted by aliens).* Furthermore, the choice of three non-key items is the most common (Cao et al., 2018). The reader is referred to Tsai (2019) for further details about the method.

Following conventional notation, let $T_i$ be the group indicator for respondent *i* where $T_i$=1 if TREATED and 0 otherwise, and $S_i$ and $R_{i,j}$ be respondent's *i* potential answers to the key item and to the *j*th non-key item (for $j = 1,2,3$), respectively. Both $S_i$ and $R_{i,j}$ are unobserved. We only observe the count of items that are true for respondent *i* ($Y_i$) so that:

$$Y_i = T_i S_i + R_i \qquad \text{where } R_i = \sum_{j=1}^{3} R_{i,j} \tag{1}$$

The prevalence of the sensitive item in the sample can be easily computed using a traditional difference-in-means estimator (DiM) as follows:

$$E(S_i) = P(S_i = 1) = \frac{\sum_{i=1}^{N} Y_i T_i}{\sum_{i=1}^{N} T_i} - \frac{\sum_{i=1}^{N} Y_i (1 - T_i)}{\sum_{i=1}^{N} (1 - T_i)} \tag{2}$$

This DiM estimator can be computed as the difference in the number of reported statements between the TREATED and the CONTROL group or as the slope of a linear regression of $Y_i$ on $T_i$.[6] Although informative, researchers are typically concerned about the different prevalence of the sensitive item among sociodemographic groups. In our context, we aim at examining how past couple break-up intentions relate to MLC. We therefore regress $Y_i$ on $T_i$, the indicator for MLC (*marlocus*), a set of control variables $X_i$, and interaction terms between the treatment indicator and *marlocus* $(T_i\ x\ marlocus_i)$, and the treatment and the controls $(T_i\ x\ X_i)$ in the following way:

$$Y_i = \alpha + \beta T_i + \gamma marlocus_i + \delta T_i \times marlocus_i + \theta X_i + \omega T_i \times X_i + \varepsilon_i \quad (3)$$

where $\varepsilon_i$ is an error term.

The model in (3) is estimated by OLS.[7] The estimates of the interaction terms can be interpreted as the heterogeneity in the prevalence of the sensitive item (past break-up intentions) based on characteristics. The estimates of the controls $(\theta)$ are expected not to be statistically significant for explaining the number of reported statements *conditional on the treatment*. That is, the TREATED and the CONTROL groups should not differ in their answers to the non-sensitive statements (no-design property).

## 5. Results

### *5.1. Descriptive statistics and diagnostic tests*

Table 5 presents the distribution of answers to the list experiment. Nobody reported the full list of statements to be true in none of the two groups. That is, none of the respondents indicated to have been abducted by aliens. This could be taken as a check that respondents understood the question. From the difference in means between the two groups, we can calculate the prevalence of past break-up intentions in the sample, which is about 46% (1.86-1.40=0.46). Nonetheless, this needs to be better explored using regression analysis.

[6] We label the indicator for being in the TREATED group (the one with the four statements including the one about break-up intentions) the 'treatment'. Note we do not formally treat those subjects with any special intervention other than the number of statements and that we cannot stablish a causal relationship but a descriptive one.

[7] Scholars have proposed alternative modelling approaches such as nonlinear-least squares and a maximum likelihood estimator that implements the EM algorithm (Imai, 2011; Blair & Imai, 2012; Imai et al., 2015). In a recent study, Ahlquist (2018) provides Monte Carlo evidence that Imai's ML estimator is heavily sensitive to small samples and relies on strong assumptions for valid implementation. The author shows that the DiM estimator requires weaker assumptions for unbiasedness and is less affected by the number of responses in the extreme tails of the distribution. Because of these reasons, in our analysis we focus on the OLS DiM estimator with interactions.

**Table 5.** Distribution of answers to the list experiment

| Number of items reported | CONTROL (N=130) | TREATED (N=136) |
|---|---|---|
| 0 | 4 | 3 |
| 1 | 70 | 41 |
| 2 | 56 | 64 |
| 3 | 0 | 28 |
| 4 | - | 0 |
| Mean | 1.40 | 1.86 |

Before that, we tested two of the three assumptions for the validity of the list experiment described before. Unfortunately, the 'no liar' assumption cannot be empirically tested. First, the treatment randomization assumption was examined by comparing the mean differences in characteristics between respondents in the TREATED and the CONTROL groups. To this end, we run a Logit regression in which the treatment status is regressed on the individual characteristics in Table 3. None of the variables is statistically significant at 95% confidence level (available upon request), which suggests that the two groups are balanced in their characteristics. To provide further evidence on this, Figure 2 presents a kernel density plot of the derived probabilities ('propensity scores') for treated and nontreated individuals. As shown, the treatment randomization condition is fulfilled.

**Figure 2.** Kernel plots of treatment propensity scores after Logit

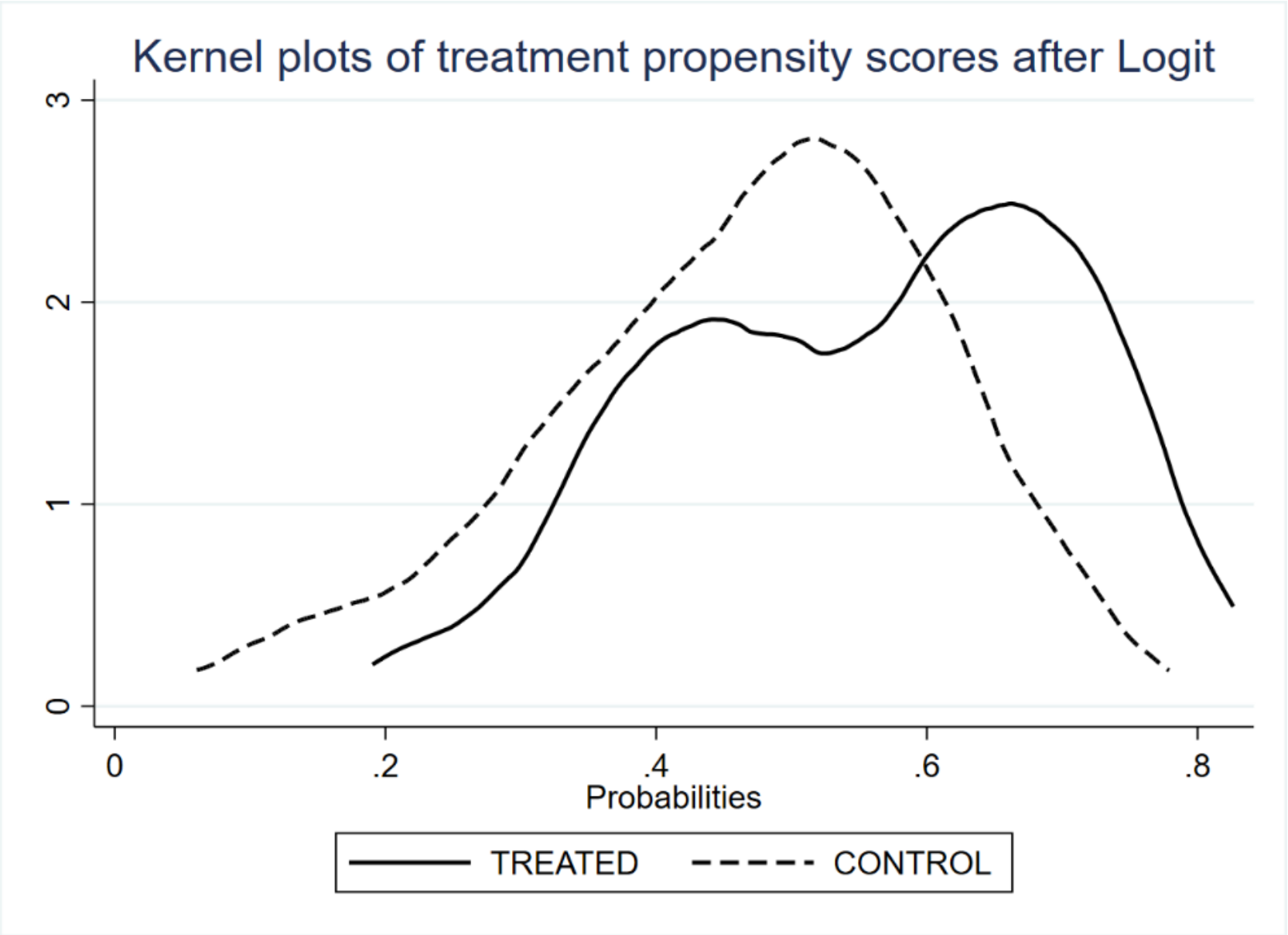

Second, the no-design hypothesis was analysed using the test proposed by Blair and Imai (2012). Intuitively, the test compares the joint probabilities of all types of item-count responses to detect whether respondents change their answers to non-key items based on being the TREATED or the CONTROL group. The test clearly suggests that the no-design assumption holds in our data (see Appendix C, Table A1). Therefore, TREATED and CONTROL groups do not differ in the sum of non-key items answered affirmatively. This validates the key identifying no-design assumption of our list experiment.

*5.2. Results from difference-in-means estimator*

Before presenting the estimation results, one important remark is in order. We acknowledge that there is scope for potential simultaneity between couple crises in the past and marital locus of control trait, which implies that *marlocus* cannot be taken as exogenous. Even though some auxiliary checks (see Online Appendix, Section D) suggest that our findings are pretty robust to moderate levels of endogeneity and locus of control has been argued to be a stable trait (Cobb-Clark & Schurer, 2013; Elkins et al., 2017), we refrain from giving our estimates a causal interpretation due to the cross-sectional nature of our dataset. Our findings are thus interpreted in terms of associations: how past intentions to separate correlate with individual marital locus of control trait. Despite we cannot establish causality (as it is unfeasible to exogenously manipulate partners' endowment of MLC trait), we nevertheless believe the net correlation between MLC and break-up intentions in the past without actual split is informative *per se*: it informs about whether beliefs about couple problems being beyond individual control actually increase the risk of dissolution.

Table 6 presents the results for the linear DiM estimator in our data. The standard errors in all the regressions are heteroskedasticity-consistent because the variance of the error term is likely to be different between the TREATED and the CONTROL groups (Blair & Imai, 2012). In column 1, we report the coefficient estimates of an OLS regression of the number of items answered affirmatively on the treatment dummy and a constant term. In line with Table 5, the prevalence of past couple break-up intentions in our sample is 46%. This coefficient can be interpreted as the *unconditional* descriptive prevalence of the sensitive item in the sample.

In column 2, we add the indicator of MLC (*marlocus*) and the interaction term between *marlocus* and the treatment dummy, which is negative and statistically significant. This means that the prevalence of the sensitive item (break-up intentions) is negatively associated with higher scores of the MLC trait. Since the sample *unconditional* prevalence of relationship instability is about 47%, a one standard deviation increase in MLC is associated with a drop in the prevalence of intra-couple problems by about 19%. In other words, the prevalence of couple break-up intentions in the past tends to be higher among those for whom relationship problems are beyond their control.

To properly isolate the linkages between marital locus of control and couple break-up intentions from other factors, in column 4 we introduce in the regression the following characteristics as controls: *female*, *age*, *univ.educ*, *income*, *numchildren, patience* and *risk*. These variables are also interacted with the treatment indicator (*T*) to see whether the prevalence of couple break-up intentions in the past varies with these characteristics.[8] Column 5 presents the average marginal effects (AME). Because there is a high collinearity between the marriage status and age (corr=0.826), in column 6 we report the results replacing age by the dummy indicator for being married.

[8] We did not include health status and nationality were not included because of lack of sufficient variability.

**Table 6.** Coefficient estimates for linear DiM regressions.

| Explanatory Variables | (1) Coefficient (SE) | (2) Coefficient (SE) | (3) AME (SE) | (4) Coefficient (SE) | (5) AME (SE) | (6) Coefficient (SE) | (7) AME (SE) |
|---|---|---|---|---|---|---|---|
| *T* | 0.460*** | 0.474*** | 0.474*** | 0.494 | 0.442*** | 0.591 | 0.444*** |
| | (0.081) | (0.080) | (0.080) | (0.441) | (0.080) | (0.362) | (0.081) |
| *marlocus* | | -0.011 | -0.107*** | -0.045 | -0.134*** | -0.016 | -0.116*** |
| | | (0.047) | (0.041) | (0.045) | (0.041) | (0.048) | (0.042) |
| *T*×*marlocus* | | -0.189** | | -0.174** | | -0.194** | |
| | | (0.082) | | (0.082) | | (0.084) | |
| *female* | | | | 0.064 | -0.030 | 0.056 | -0.036 |
| | | | | (0.102) | (0.083) | (0.109) | (0.084) |
| *T*×*female* | | | | -0.183 | | -0.180 | |
| | | | | (0.165) | | (0.168) | |
| *age* | | | | -1.601*** | -1.505*** | | |
| | | | | (0.564) | (0.524) | | |
| *T*×*age* | | | | 0.189 | | | |
| | | | | (1.039) | | | |
| *married* | | | | | | -0.198 | -0.119 |
| | | | | | | (0.168) | (0.172) |
| *T*×*married* | | | | | | 0.154 | |
| | | | | | | (0.341) | |
| *univ.educ* | | | | -0.068 | 0.130 | -0.016 | 0.160* |
| | | | | (0.104) | (0.089) | (0.105) | (0.091) |
| *T*×*univ.educ* | | | | 0.387** | | 0.344* | |
| | | | | (0.178) | | (0.181) | |
| *income* | | | | 0.046 | 0.046 | -0.002 | 0.006 |
| | | | | (0.050) | (0.037) | (0.047) | (0.035) |
| *T*×*income* | | | | -0.001 | | 0.014 | |
| | | | | (0.074) | | (0.069) | |
| *numchildren* | | | | 0.158* | 0.009 | 0.066 | -0.106 |
| | | | | (0.094) | (0.080) | (0.089) | (0.091) |
| *T*×*numchildren* | | | | -0.291* | | -0.337* | |
| | | | | (0.160) | | (0.179) | |
| *patience* | | | | 0.007 | -0.005 | 0.004 | -0.007 |
| | | | | (0.019) | (0.016) | (0.019) | (0.016) |
| *T*×*patience* | | | | -0.022 | | -0.022 | |
| | | | | (0.032) | | (0.032) | |
| *risk* | | | | -0.015 | -0.014 | -0.000 | -0.005 |
| | | | | (0.017) | (0.018) | (0.017) | (0.018) |
| *T*×*risk* | | | | 0.002 | | -0.009 | |
| | | | | (0.035) | | (0.035) | |
| constant | 1.400*** | 1.399*** | | 1.830*** | | 1.392*** | |
| | (0.048) | (0.048) | | (0.241) | | (0.207) | |
| Observations | 266 | 266 | | 266 | | 266 | |

Robust standard errors in parentheses. *** p<0.01, ** p<0.05, * p<0.1

We find that the interaction term between *marlocus* and the dummy indicator for being part of the treatment group is consistently negative and significant when adding the full set of controls. Interestingly, the interaction between university education and the treatment dummy is also negative and significant. This indicates that, conditional on MLC and the other controls, educated individuals are significantly more likely to have seriously considered the possibility of breaking with their current partners in the past. We also document that the number of items answered positively is positively associated with age. Although we guarantee treatment randomization, this result may reflect the fact that the likelihood of "having celebrated a meeting with my family or some friends in the last month" (item b) could be lower for older people (in our sample). Importantly, none of the remaining control variables are significant. Note that this non-significance is a desired feature, since the identification of the prevalence of the sensitive item needs the TREATED and the CONTROL groups to be comparable. A relevant consequence of this is that couple break-up intentions are not correlated with our indicators of patience and risk attitude.

Please note that the non-significance of the treatment dummy alone in the expanded regressions is due to being interacted with all the covariates. The *conditional* descriptive prevalence of break-up intentions ($\tau$) in columns 5 and 7 is 0.44. This value is close to the simplest DiM estimator in column 1 without any control, but slightly lower. Therefore, the prevalence of past couple instability is robust also in the regressions controlling for observable sources of heterogeneity. Because the marginal effects in the model with interactions depend on the values of all the explanatory variables, we derived the estimate of $\tau$ at different values of the control variables (based on the estimates from column 4). This is equivalent to compute $\tau$ for different subsamples but based on the model predictions from the pooled data. The corresponding values are presented in Table 7.

**Table 7**. Estimates of the descriptive prevalence of sensitive item by subgroups

| | $\tau$ |
|---|---|
| Males (*females*=0) | 0.466*** |
| Females (*females*=1) | 0.418*** |
| University studies (*univ.educ*=1) | 0.567*** |
| No children (*numchildren*=0) | 0.552*** |
| No income (*income*=0) | 0.472*** |
| Low income (*income*=1) | 0.467*** |
| Middle income (*income*=2) | 0.467*** |
| Middle-high income (*income*=3) | 0.406*** |
| High income (*income*=4) | 0.229 |
| 1st quartile MLC (*marlocus*=-0.755) | 0.573*** |
| Median MLC (*marlocus*=-0.035) | 0.448*** |
| 3rd quartile MLC (*marlocus*=0.683) | 0.323*** |

Note: the figures are calculating as average effects by subgroups based on the estimates in column 4 in Table 6.
*** p<0.01, ** p<0.05, * p<0.1

Interestingly, past couple break-up intentions are relatively higher among males, those with university studies and couples without children. These results are in line with Becker et al. (1977) and Becker (1974). Couples with children are more committed to each other because such couple-specific capital worth less in any other relationship or when turning back to singlehood. Therefore, children tend to serve as a commitment device. If we focus on the role of education, we note that although high-educated individuals are expected to gain more from the relationship due to having greater market and nonmarket skills, at the same time their expected utility of engaging in any other relationship is also larger. Assuming that high levels of education are a desired feature in the remarriage market, these individuals have better outside opportunities. Concerning gender differences in the prevalence of couple dissolution intentions, females are more likely to conform to social norms (Soons & Kalmijn, 2009) and have been shown to be more committed than males (Kwang et al., 2013). Moreover, since they are typically the secondary wage earnings, they have lower exit possibilities (Parkman, 2004). We also document that the prevalence of break-up intentions exhibits a negative relationship with income. This could be explained by the fact that stress and strain are more likely to emerge under financial constraints (Britt & Huston, 2012), especially if it is the male who is at a greater risk to become unemployed (Ahituv & Lerman, 2011).

The likelihood of past couple instability is negatively correlated with higher scores of *marlocus*. The prevalence of break-up intentions for individuals in the third quartile of the MLC distribution is only 32%, but it amounts to 57% for those in the first quartile. Again, couple instability seems to be greater among those who consider intra-couple problems to be due to chance and who devote lower effort towards searching for Pareto-improving solutions.

### *5.3. Robustness checks*

We have performed several robustness checks and the estimation results are presented in Appendix C. First, because our dependent variable is a count, we have estimated all the regressions using a Poisson model. All the results are consistent. Second, we replace the individual indicator of MLC (*marlocus*) by the male-female and absolute value difference in MLC between partners, respectively. None of the interaction terms are statistically significant, suggesting that break-up intentions are not affected by the gap in MLC between spouses but by the individual endowment of MLC. Third, we constructed the index for MLC based on Principal Component Analysis with one component. The Kaiser-Meyer-Olkin measure of sampling adequacy is 0.875. When we repeat the regression analysis using this variable in place of *marlocus,* we get very similar results. Therefore, the findings appear not to be affected by the index construction. Fourth, we repeated our analysis using clustered standard errors at the couple level to control for potential cross-sectional dependence. The statistical significance of the coefficients remains also unchanged. Finally, we have introduced other control variables such as the number of working hours per week, self-reported couple satisfaction or indicators describing the Big Five personality traits.[9] These variables were introduced alone and interacted with the treatment dummy. None of them have significant coefficient estimates while the coefficient estimate for *T*×*marlocus* remains largely unchanged, so we excluded them from the specification to save degrees of freedom (available upon request).

## 6. Conclusions

The main goal of this paper has been to examine the relationship between past serious intentions to end the relationship and marital locus of control, a trait that measures individual's perception of control over couple problems. Because answers from direct asking could be affected by social desirability bias, we have implemented a list experiment to elicit individual's break-up intentions in the past in an indirect way. The data has been collected as a part of a lab experiment involving current married and non-married couples living in four cities in the north of Spain. The analysis of couple instability in urban areas is particularly relevant also because matching theories show that those living in cities face lower search costs and higher access to potential partners.

[9] The indices describing the Openness, Conscientiousness, Extraversion, Agreeableness, and Neuroticism were derived from the ten-item personality inventory (TIPI) developed by Gosling et al. (2003).

Using a linear difference-in-mean estimator with interactions, we have found that about 44% of our sample has seriously considered to end their relationship at least once in the past. As hypothesized, we have shown that the likelihood of past intra-couple problems is negatively associated with high scores in marital locus of control. This implies that those who perceive that their actions have little influence on couple problem solving are significantly more likely to have been about to finish their partnership at least once in the past. The mechanism explaining this pattern is that less serious couple problems might occur among partners who consider they have personal control over intra-couple disagreements and the capacity to discuss and solve them. Additionally, our results indicate that couple instability in the past is larger among males, university graduates and couples without children. In contrast, we have shown that break-up intentions are not correlated with patience and the willingness to take risks. Therefore, the influence of marital locus of control on break-up intentions appears not to be confounded by these other traits. Patience and courage are soft skills that could be expected to correlate with MLC and, in general, a person's predisposition to solve conflicts.

The paper contributes to the economic and sociological literatures on couple instability in several regards. In line with Lundberg (2012), we have documented that attitudes are important in maintaining positive relationship surplus and stability. Notably, by studying past break-up intentions in a sample of current couples, our analysis has offered a different approach to the analysis of a couple functioning. To the best of our knowledge, this is the first study that addresses past attempts to separate among couples that are currently together and relates strains in the relationship to the marital locus of control trait. Accordingly, we shed light on the role of problem-solving attitudes in conflicts resolution within the couple.

Our study has important policy implications. Over the course of a sentimental relationship, either with or without formal marriage, disagreements often emerge. The ability to sustain the relationship depends on the self-perceived capacity to manage crises during a relationship. If MLC is so relevant in preventing family break-ups, then the question of social interest is how to "produce" the soft skill internal locus of control in the new generations. This question is waiting for extra research efforts. Regarding the existing generations, instead, it seems very important to invest more public resources on family enrichment programs that focus on strengthening the individual endowment of marital locus of control.

Interventions designed to enhance individual's sense of control and capacity of proactively dealing with couple problems would be beneficial to improve couple commitment. This is in line with the literature on family enrichment in the social psychology literature (Neal et al., 2014). Couple therapists need to further enhance individual's locus of control over couple difficulties to improve their capacity and attitude towards problem-solving. Preventive programs aimed at training partners' coping skills in terms of making the other feeling understood and listened could be highly beneficial, not only for the couple sphere but also for other domains of life. This is further robust evidence about the social and economic relevance of the many non-market benefits associated with public investments in the formation of soft skills and non-cognitive abilities.

As part of our research agenda, we intend to extend the analysis to a larger sample placing special emphasis on gender roles, the type of conjugal relation, the gender division of household chores and social norms. Besides, togetherness has been recently revealed as an important gain from sentimental relationships (Cosaert et al., 2023). It would be interesting to explore the moderating effects of spending time together for explaining the connection between marital locus of control trait couple stability.